# Investigating SCADA Failures in Interdependent Critical Infrastructure Systems

Razgar Ebrahimy
School of Computing Science, Newcastle University,
Newcastle Upon Tyne, UK
Razgar.ebrahimy@ncl.ac.uk

*Abstract*— this paper is based on the initial ideas of a research proposal which will investigate SCADA failures in physical infrastructure systems. The results will be used to develop a new notation to help risk assessment using dependable computing concepts. SCADA systems are widely used within critical infrastructures to perform system controls and deliver services to linked and dependent systems. Failures in SCADA systems will be investigated to help understand and prevent cascading failures in future.

## I. Introduction

Critical Infrastructures are those kinds of systems and assets, physical or virtual, so important and vital to nations and societies that destruction and failure of them could have a debilitating impact on national security, economy and national health and safety [1], [2]. Critical and national infrastructures form the backbone of our modern communities [3] and they are the fundamental factors in providing economic productivity and social wellbeing [4].

Physical infrastructure consists of energy, transport, waste, water and the newly emerged ICT sector. Moreover very high cross-sector interdependency exists between all these sectors [6]. These interdependencies highlight the importance of availability of services by each sector in order to provide services and facilitate other dependent sectors. Generally, failures of one infrastructure could lead to the failure and stoppage of services in another dependent infrastructure. For instance, all infrastructures rely on availability of electricity, hence power outage can have direct impact on availability of other services, such as transport, emergency, finance, communication, water supply etc.

Infrastructure assets have evolved over a period of time. Moreover infrastructure projects developments and maintenance are very costly for governments to fund [37]. As a result modeling and simulation tools are used to aid the policy makers in planning and investing in infrastructure. Some of these tools can be very helpful and used to assess risks, opportunities and threats analysis. Some of these tools can indicate what the existing infrastructure could face in future, for instance analyzing supply and demand, investment and risk analysis. To show the role and importance of these modeling tools, a dedicated section describes them in more detail.

The research reported in this paper was part of the UK Infrastructure Transitions Research Consortium (ITRC) funded by the Engineering and Physical Sciences Research Council under Programme Grant EP/I01344X/.

The UK Infrastructure Transitions Research Consortium (ITRC), which was established jointly by several UK universities, aims to inform the analysis, planning and design of national infrastructure through the development and demonstration of new decision support tools, and working with partners in government and industry. This research is part of an ITRC project with focus on risks and failures of ICT infrastructure.

ITRC's initial interest in ICT has been to analyze the development and upgrading of current communication systems in the UK infrastructure. Another objective has been to investigate ICT impacts, including failures and opportunities to other sectors, and also to quantify and measure the values that the ICT sector and component add to the operation of other infrastructure systems. As a result a fast track analysis report about ICT constraints [38] has been produced that investigates the current state of ICT in the UK and mostly focuses on communication infrastructure. However this research aims to extend this idea of focusing on telecommunication and instead investigates ICT in more detail by looking into software failures and limitations in infrastructure systems. This is to analyze how failures and limitations of some of these ICT Systems could impact and influence the infrastructure sectors. Interdependency interaction analysis between different sectors of national infrastructures is a core part of the ITRC project and this research.

Dependable computing concepts [7]-[8] are the framework for analyzing and identifying ICT failures in this research.

This research aims to investigate in more detail by identifying ICT failures and threats in infrastructure systems and use these understandings to develop a notation which could then be used for risk assessment, future development and disaster recovery procedures and modeling tools.

Some of the infrastructure systems are old and not designed to handle interdependency, however it is not feasible to upgrade all computing software in infrastructure at once, mainly because of the cost. As a result we face some sort of software limitation when merging new computing systems with old systems. This can lead to new and unknown vulnerabilities in overall systems. In addition some of these new complex systems, such as smart meters [43] that are to be deployed in the UK, could face new technical, ethical and security challenges.

The following sections of this paper include background analysis by investigating interdependency, modeling techniques, dependability concepts and cyber threats. Problems and justification are covered in Section III and finally the high level presentation of timetable is presented in Section IV.

## II. Background and related work

The main aim of this research project is to investigate and analyze the ICT failures in physical infrastructure and specifically how these failures could propagate through these interdependent sectors. Infrastructure sectors are known to be a combination of 'systems of systems', that are all directly or indirectly connected and dependent on each other [1].

There have been a number of studies about complex systems and interdependent networks [1] with the objective to design more robust and reliable systems and networks. Although some of these studies are not directly relevant to aging infrastructure, their general methodologies and approaches such as Top-down or Bottom-up approaches can be implemented in design, planning and analyzing the existing infrastructures.

This section looks into the concept of interdependency in infrastructure including direct and indirect dependencies and also the strength and vulnerabilities of such interconnected systems. It also analyzes some of the infrastructure modeling techniques used to model fragments of some of these sectors, and shows how helpful these models could be in planning and forecasting future development and failure preventions.

The section also looks into dependable computing concepts [7]-[8] and how they could be related to infrastructure systems. Finally this section looks at the cyber interdependency of these sectors and vulnerabilities and threats that modern infrastructure faces.

### A. Infrastructure as complex and interdependent systems

In general most studies have focused on a single and isolated sector in order to get better understanding of failures and operations of these systems [10]-[11]. However infrastructures share similar attributes as they are all complex adaptive systems [9], and they show a large number of interdependencies in different ways [1] as follows:

- **Physical interdependency** when energy, material or people migrate from one infrastructure to another.
- **Cyber interdependency** is when information is transmitted or exchanged between infrastructures. For instance, the reliance of some ICS (Industry control systems) in the energy sector on the transmission of data over networks for controlling purposes or traffic data in the case of the transport sector.
- **Geographical interdependency** is a close spatial proximity of the elements of infrastructure such that a failure in a node in an area could have an effect on the other nearby infrastructures. For instance flooding in an area where the road for transporting coal to a power station is blocked could result in power outage and cascade through the neighbouring nodes.
- **Logical interdependencies** that can be a combination of all other types of connections such as financial dependence, political coordination and so on.

Advances in cyber-based systems and communication infrastructure have resulted in increased coupling of entire infrastructures to become more robust and efficient. Nevertheless the close coupling introduces new vulnerabilities in these dependent systems [13] such as cascading failures.

One of the key points and properties of interdependent infrastructures is that sometimes even a minor failure can lead to failure of dependent infrastructure and cascade to other networks recursively. For instance the 2003 blackout in which large parts of Midwest and Northeast United States and Ontario, Canada experienced electrical power outage. Another example of cascading failure is the 2003 electrical blackout that affected much of Italy. In this second case the power outage resulted in failure of Internet communicating nodes, which in turn caused more power station breakdowns [12].

Some of these interdependencies are straightforward, easy to analyze and investigate. For example all infrastructures rely on availability of electricity to operate. However it becomes more sophisticated and complicated once we begin to look at the cyber dependency failures of these systems and to show the impacts on all dependent systems. While some of this interdependency is clear and easier to identify, the rest might be unclear and can only be understood with complex system analysis [14]-[15].

With such interdependencies in infrastructure the terms failure and fault have different meanings based on each sector. For instance material fault and software fault are two different notations with different origins and implications. The first one might be caused by natural hazard or usage whereas the second one could be as a result of human error [16].

Infrastructures are complex in their own right and not easy to analyze, specifically when considering the market, government regulations, economic benefits, policy making and technical aspects that need to be considered for each sector. It is also true that infrastructures do not exist individually and are highly interdependent. To fully understand the scope and operational characteristic of infrastructures their interdependencies must be integral to any analysis [17].

### B. Modeling and simulation techniques

Modeling and simulation are known to be essential for ensuring the safe, reliable and continuous operation of critical infrastructure [17]. There are a number of modeling techniques and tools available to analyze the individual and interdependent sectors, each of them to analyze different aspects of infrastructure. There are various well developed commercial models and simulation tools of individual infrastructure available to help the owners to operate and manage systems.

Some of these modeling techniques can be separated into six different categories as follows [17]-[21]-[20]:
- **Agent Based Model (ABM)** is a class of computational models for simulating the actions and interactions of autonomous agents with a view to assessing their effects on the system as a whole. ABMs have been used widely in infrastructure and interdependency analysis [18]-[19]. ABMs are capable of capturing interactions in an accurate, yet simple, fashion. By using ABM it is possible to analyze the characteristics and state of an infrastructure by modeling the physical component as an agent.
- **Physical Based Models (PBM)** can be modeled and analyzed with standard engineering techniques. For instance power flow and stability analysis can be performed on the electric power grids. This analysis can go into more detail at component level and examine issues like power outage with single or multiple associated connected nodes [20].
- **Population Mobility Models**: The examination of entities through urban regions and their interaction with each other.
- **Economic Models**: Including Leontief input-output models of economic flow.
- **Dynamic Simulations**: Modeling the generation, distribution and consumption of infrastructure commodities and services as flows and accumulations.
- **Aggregate Supply and Demand Tools**: An evaluation of the total demand for infrastructure services in a region together with the ability to supply those services.

There have been attempts to model Cyber-Physical Systems (CPS) using semantic agents [22]; however it is not possible to accurately model and represent these systems in a way that covers all necessary operations of interdependent sectors using CPSs [23].

In addition to modeling tools and techniques, simulation frameworks that allow coupling of multiple interdependent infrastructures have emerged [5]. For instance the National Infrastructure Simulation and Analysis Centre (NISAC) based in USA has developed tools to address cyber and physical dependencies and interdependencies in an all-hazards context [5]- [24].

There is a wide range of literature available which investigates the interdependency, risk analysis and supply and demand chain. However there is lack of models and simulation tools which consider the ICT sector in connection with other infrastructures, and are capable of analyzing ICT failures or predicting future behaviors.

*C. Dependable Systems*

The notation of dependability is the property of a computer system that allows reliance to be justifiably placed on the service it delivers [7]. A dependable system should have the following attribute: availability, reliability, safety, integrity and maintainability [8].

Based on [7] an error could lead to failure if there is lack of system compositions such as redundancy or the definition of failure from the user's viewpoint. For instance [25] argues that scheduled maintenance of a power plant which needs to be shut down and does not provide service should not be considered as a system failure. In the dependability concept [7-8] fault, error and failures are described as follow:
- **Fault** is the cause of an error.
- **Error** is part of the system state which could lead a service failure
- **Failure** occurs when the delivered service deviates from the correct service.

Complex infrastructure systems and complex software systems both share some similarities such that failures in each of these systems can have direct or indirect effects on their dependent systems.
- **Cascading failure** is a disruption in one system that could cause a fault in another system. For example in infrastructure we can refer back to the 2003 blackout in America that led to communication and water supply outage and air traffic disruption [26]. The cause was a software bug in the alarm system at a control room of the FirstEnergy Corporation in Ohio [26].
- **Escalating failure** is a disruption in one infrastructure that exacerbates an independent disruption in a second infrastructure. For instance the recovery and restoration of service in one infrastructure is made more difficult because another infrastructure is not available.
- **Common cause failure** is disruption of two or more infrastructures at the same time as the result of the same cause. For instance flooding could impact electric power, transport system, communication, gas and water supply.

Dependability concepts are specifically helpful in this research, as the ICT failures in infrastructure will be measured against these dependability notations.

*D. Cyber threats to infrastructure*

Infrastructures are highly dependent on ICT. The type of dependency varies significantly depending on the purpose and service of the ICT component. For example, supervisory control and data acquisition systems (SCADAs) are part of operational networks, composed of computer systems that yield the operational ability to supervise acquire data from and control the physical process [27]. Classically SCADA systems were not designed to be widely distributed or remotely accessible. However most of these systems are now connected to the Internet and could be vulnerable to dependency failures and malicious attacks [28]-[29].

For instance the vulnerabilities of a power system include three main components: computer, communication, and the power generator itself [39]. Attacks can be targeted at specific systems, subsystems, and multiple locations simultaneously from a remote location. These types of vulnerabilities are all as a result of cyber dependency.

One of most famous infrastructure threats has been the Stuxnet virus, which was designed with painstaking precision to burrow deep into Iran's nuclear program and destroy physical infrastructure [41]. This is an indication to show how serious and devastating an ICT component failure could be for national infrastructure. Interestingly even after Stuxnet was detected another far less sophisticated malware called Mahdi managed to compromise engineering firms, government agencies and financial services firms in the Middle East [42]. his is an indication that there is still not a clear understanding of how these ICT systems could be so vulnerable to even low level attacks.

## III. Problems and Justifications

Infrastructure is fundamental to the functioning of economy and our daily lives. It is important that infrastructure is available and reliable at all times. Infrastructure has been developed over decades and in some cases even centuries. Therefore it is not possible and cost effective to try to redesign and reinvent everything from scratch. In 2010 UK engineering works were valued at £790 billion [30]. As discussed in Section II infrastructure is highly interdependent. However there are risks and opportunities associated with such interdependencies. ICT can bring opportunities to other sectors and society by providing ease of functionality and services, such as smart metering [31], providing faster communication and more technological improvements and new inventions in providing services to other infrastructure systems such as safety systems in the transport sector.

However with such rapid integration and dependency on ICT, the current infrastructure systems are becoming even more complex to analyze and model. Complex systems can exhibit surprising behavior due to their complexity [32]. Current complex systems show an increase of structural, dynamic, functional and algorithmic complexity [32]. Such system complexities can cause challenges in design, operation, reliability and efficiency.

Most of the focus on the ICT sector has been about communication and facilitating other infrastructures [4], for instance ensuring the availability of broadband and network services and upgrading current UK communication infrastructure by providing fiber optic broadband to consumers and other sectors [33]. By considering communication as a component of ICT it is easier to understand and to analyze availability, improvement and to measure failure impacts of the communication sector. However once we try to look at the ICT sector as a whole and take a top-down approach to investigate the impact failures, we begin to face uncertainty due to the complexity of the ICT sector, lack of data and diverse ownership of the sector.

One of the main objectives of this research is to investigate beyond telecommunication and understand failures in more detail by looking at operational software systems such as SCADAs, process control systems (PCS) or distributed control systems (DCS), which is a complex combination of SCADA and PCS. To do this it is important to categorize types of failures and their possible impacts on infrastructure. Using these types of failures we can build a bigger picture of the connected systems and predict future behaviors. This is specifically helpful in avoiding cascading failures, which could be the result of an ICT component failure in infrastructure. Generally ICT and specifically SCADA systems all individually rely on software development procedures and offer some degree of fault tolerance within their design [34]. However since most of these systems are now connected to internal and external operational systems therefore they can be vulnerable to new types of failures that have not been considered when they were designed [27].

Using these investigation results we can then model ICT components in infrastructure. Although it is very complicated and expensive to model and simulate the ICT sector as a whole with its interdependencies to other sectors, if we know of the factors that cause failures we might be able to measure and model some portions of ICT failures with some degree of interdependencies which could be helpful in future risk assessment, planning and improvements of existing infrastructure.

There are a number of alternative investigation methods available, such as top-down or bottom-up approaches. A top-down approach in conjunction with formal methods [35] is one of the options as it can be used to eliminate the details and instead focus on the overall system. However the downside is that we are dealing with a combination of existing infrastructures. This method works best if we plan and design a new infrastructure system or component from scratch as it can lead us to think about all current and future implications. With such complexity in infrastructure systems it is not feasible and cost effective to redesign everything, however the idea of a top-down approach could be used in cases of new developments such as hospitals [36], new railway systems or roads.

Some researchers believe that when complex systems reach a certain size of complexity then algorithmic constraints often prohibit efficient top-down management and design [32]. It is suggested that self-organization is an alternative idea to manage complex systems. In this research we use a combination of the two approaches. For instance we could use the top-down approach to identify the possible failures and use the bottom-up approach to decentralized the component at a micro level and allow the system component to either self-organize or avoid cascading failures by having some kind of redundancy in place.

Cyber threats to critical infrastructure and SCADA failures could be used as case studies for this research since they both cover interdependencies and can have direct impact on overall operation of infrastructure. Because of the nature of this type of failure, the availability of data can be very limited. If, however, micro level failures are considered in each case (using the developed notation) this research could produce outcomes to show the extent and implications of each type failure on national infrastructure systems.

## IV. Conclusion

One of the key points in this research is to draw a clear understanding and identification of ICT failures in micro details. Then depending on the types of failures, they can be categorized in different orders and magnitudes of impact. For instance a computer server failure in a power station will have different impacts than the failure of a sensor that triggers when an electric fuse in an electric box fails within the same power station. The server failure might shut down the station whereas fuse failure can only impact a small area. Identifying these types of failures at the micro level will help in risk assessment and strategy planning and recovery procedures. Such failures can also be used in modeling some parts of the ICT sector.